\documentclass[useAMS,usenatbib]{mn2e}
\usepackage{graphicx} 
\usepackage{grffile}
\usepackage{longtable}
\usepackage{rotating}
\title[The VMC Survey. VIII. First results for Anomalous Cepheids]{The
  VMC Survey. VIII. First results for Anomalous Cepheids\thanks{Based on
    observations made with VISTA at ESO under programme ID 179.B-2003.}}
\author[V. Ripepi et al.]{V. Ripepi$^{1}$\thanks{E-mail:
ripepi@oacn.inaf.it},
M. Marconi$^{1}$, 
M. I. Moretti$^{1}$, 
G. Clementini$^{2}$,
M-R. L. Cioni$^{3,4}$,
\and
R. de Grijs$^{5,6}$, 
J. P. Emerson$^{7}$
M.A.T. Groenewegen$^{8}$,
V. D. Ivanov$^{9}$,
J. M. Oliveira$^{10}$
\\
$^{1}$ INAF-Osservatorio Astronomico di Capodimonte, Via Moiariello
 16, 80131, Naples, Italy \\
$^{2}$ INAF-Osservatorio Astronomico di Bologna, via Ranzani 1,
Bologna, Italy \\ 
$^{3}$ University of Hertfordshire, Physics Astronomy and
 Mathematics, Hatfield AL10 9AB, UK \\
$^{4}$ Leibnitz-Institut f\"{u}r Astrophysik Potsdam, An der
Sternwarte 16, 14482 Potsdam, Germany; \\
Research Fellow of the Alexander von Humboldt Foundation\\ 
$^{5}$ Kavli Institute for Astronomy and Astrophysics, Peking
University, Yi He Yuan Lu 5, Hai Dian District, Beijing 100871, China \\
$^{6}$ Department of Astronomy, Peking University, Yi He Yuan Lu 5, Hai Dian District, Beijing 100871, China\\
$^{7}$ Astronomy Unit, School of Physics \& Astronomy, Queen Mary
University of London, Mile End Road, London E1 4NS, UK\\
$^{8}$ Koninklijke Sterrenwacht van Belgi\"e, Ringlaan 3, 1180, Brussel, Belgium\\
$^{9}$ European Southern Observatory, Ave. Alonso de Cordova 3107, Casilla 19, Chile \\
$^{10}$ Astrophysics Group, Lennard-Jones Laboratories, Keele
University, Staffordshire ST5 5BG, UK\\
}
\begin{document}

\date{}

\pagerange{\pageref{firstpage}--\pageref{lastpage}} \pubyear{2002}

\maketitle

\label{firstpage}

\begin{abstract}
  The VISTA near-infrared $YJK_\mathrm{s}$ survey of the Magellanic Clouds System (VMC, PI M.-R. L. Cioni) is collecting
  deep $K_\mathrm{s}$--band time--series photometry of the pulsating
  variable stars hosted in the system formed by the two Magellanic
  Clouds and the Bridge connecting them.  In this paper 
we present for the first time $K_\mathrm{s}$--band  light curves for 
Anomalous Cepheid (AC) variables.  
In particular, we have analysed a sample of 48  Large
Magellanic Cloud ACs, 
for which  identification  and  optical magnitudes were 
obtained from the OGLE~III and IV catalogues.
The VMC  $K_\mathrm{s}$--band  light curves for ACs  are well sampled,
 with the number of epochs ranging 
from 8 to 16, and allowing us to obtain very precise mean $K_\mathrm{s}$ magnitudes with 
errors on average of the order of 0.01 mag. 
The $\langle  K_\mathrm{s} \rangle$ values were used to build the first Period--Luminosity and
Period--Wesenheit relations in the near-infrared for fundamental-mode and
first overtone ACs.  
At the same time we exploited the optical
($V,I$) OGLE data to build accurate Period--Luminosity,
Period--Luminosity--Colour  and Period-Wesenheit relations both for
fundamental-mode and first overtone ACs. For the first time these relations were derived from 
 a sample of pulsators which uniformly cover the whole AC instability strip. 
The application of the optical Period--Wesenheit relation to a sample of dwarf galaxies hosting 
a significant population of ACs
revealed that this relation is a valuable tool for deriving distances
within the Local Group. Due to its lower dispersion, we expect 
the $K_\mathrm{s}$ Period--Wesenheit relations first derived in this paper to represent a valuable 
tool for measuring accurate distances to galaxies hosting
ACs when more data in near-infrared filters become available.
\end{abstract}


\section{Introduction}

The Magellanic Clouds (MCs) play a fundamental role in the context of 
stellar populations and galactic evolution  studies \citep[see
e.g.][]{HZ04,HZ09}. Together with the Milky Way they form the 
 closest example of ongoing complex interaction among galaxies
 \citep[see e.g.][]{putman98,muller04,stanim04,bekki07,Venzmer2012,For2013}. Moreover,
 since they are more metal poor than our Galaxy and host a
 significant population of younger  populous clusters, the
 MCs represent an ideal laboratory for testing physical and
 numerical assumptions of stellar evolution codes \citep[see e.g.][]{Matteucci2002,bro04,nl12}.

In the framework of the extragalactic distance scale, the Large
Magellanic Cloud (LMC) represents
the first crucial step on which the calibration of Classical Cepheid (CC) 
Period-Luminosity ($P$--$L$) relations and in turn of secondary
distance indicators relies \citep[see e.g.][and references therein]{f01,w12,Rie11,degrijs2011}.
Similarly, the LMC hosts several  thousand  RR Lyrae variables,
which represent  the most important Population II standard
candles  through the well known $M_V(\rm RR)$--[Fe/H]  and 
near-infrared (NIR) metal dependent $P$--$L$ relations.  
Hence, the LMC is the ideal place 
to compare the distance scales derived from Population I and
II indicators \citep[see e.g.][and references therein]{Clementini2003,w12}.


In this context NIR observations of  pulsating
stars \citep[see e.g.][and references therein]{Ripepi12a,Moretti13} are very promising.
Such observations 
over the whole Magellanic system, including the relatively unexplored Bridge connecting
the two Clouds, represent one of the most important aims of the {\it VISTA
   near-infrared $YJK_\mathrm{s}$ survey of the Magellanic Clouds system} 
\citep[VMC;
][hereinafter Paper I]{Cioni11}. The VMC ESO public survey is acquiring
deep NIR photometric data in the $Y$, $J$ and $K_\mathrm{s}$ filters
on a wide area across the Magellanic system, with the VIRCAM camera
\citep{Dalton_etal06} of the ESO VISTA telescope
\citep{Emerson_etal06}.  The principal scientific aims of VMC are the
reconstruction of the spatially-resolved star-formation history (SFH)
and the determination of the 3D structure of the whole Magellanic
system. The observational strategy, planned to go as deep as $K_\mathrm{s} \sim 20.3$
mag (Vega) at Signal to Noise ratio (S/N)=10, will allow us to detect sources 
encompassing most phases of
stellar evolution, namely the main-sequence, the subgiant branch, the upper and
lower red giant branch (RGB), the red clump, the RR Lyrae and
Cepheid location, the asymptotic giant branch (AGB), the post-AGB and
planetary nebulae (PNe) phases,  but also supernova remnants (SNRs), etc. These
different stellar populations will allow us to study  age and
metallicity distributions within the whole MC system.

The properties of the CCs and RR Lyrae stars observed by
the VMC survey are addressed by
\citet{Ripepi12a,Ripepi12b}. In these papers, 
the authors provide important results on the calibration of the 
distance scales for both these important standard candles.
Beyond these two classes of variables, other kinds of pulsating stars play an
important role both as distance indicators and stellar population
tracers. In particular, Anomalous Cepheids (ACs) are Pop.II stars, according to
their low metallicity,  with periods shorter than 2 days and brighter
than RR Lyrae stars by an amount that typically ranges from 0.3 mag
for the shortest period  to 2 mag for the longest period pulsators 
\citep[see e.g.][and references therein]{c98,mfc04,c04}.
From the evolutionary point of view these variables are usually associated with
the central He burning phase of stars with masses from $\sim$1.3 to
$\sim$2.1 $M_{\odot}$ and metallicities lower than $Z=0.0004$ ([Fe/H]$\approx$$-$1.7 dex, for $Z_{\odot}=0.02$) 
\citep[see][for details]{c98,mfc04}.
In this low metallicity regime the Zero Age Horizontal Branch (ZAHB) is
predicted to show a turnover at lower effective
temperatures. Indeed, as the mass increases along the ZAHB, the
effective temperature decreases down to a minimum value beyond which
both the effective temperature and the luminosity increase up to the
values corresponding to the transition mass limit \citep[see e.g.][for
details]{c98}. Above this limit,
the star ignites the triple--$\alpha$ reaction quiescently and burns central
Helium along the blue loop phase, becoming a CC when
crossing the instability strip.
The location of the ACs in a period magnitude plane is therefore the
downward extension of the distribution of metal poor classical
Cepheids, and in the short period range the discrimination between the
two classes can be risky, as extensively discussed in  \citet[][]{c04}.
These authors have also demonstrated that both the predicted limiting
magnitude for massive pulsators and the magnitude of RR
Lyrae stars, depend on the assumed metallicity, with the
difference between these magnitudes increasing with the metal content.

In the LMC there are 86 ACs \citep{sos08,sos12}. 
An investigation of the properties of these pulsators in the optical
bands has been recently presented by \citet{fiomon12}, using these
objects as stellar population tracers.
These authors derived pulsation constraints on their mode identification and
mass estimate. Nevertheless, optical Period--Luminosity--Colour
($P$--$L$--$C$) and Wesenheit ($P$--$W$) relations \citep[see][for a
detailed discussion]{m82,mf91} for the LMC ACs
are still lacking in the literature. These relations are important tools 
providing an alternative and independent estimate of the distance to the LMC. Furthermore, 
for the reasons discussed
above, we expect significant advantages when deriving similar relations
in the NIR bands. 
\\

The VMC data for the ACs are presented in Section 2. The $P$--$L$, $P$--$W$, and
$P$--$L$--$C$ relations in the optical (on the basis of the OGLE data) and
in the NIR (on the basis of the VMC data) derived for
Fundamental-mode (F) and
First Overtone (FO) ACs are discussed in Sections 3 and 4.
The zero-point calibrations of the AC  $P$--$L$,
$P$--$W$, and $P$--$L$--$C$ relations and their application are
presented in Section 5.
Finally, a summary of the main  results is presented in Section 6.

\begin{figure}
\includegraphics[width=8.8cm]{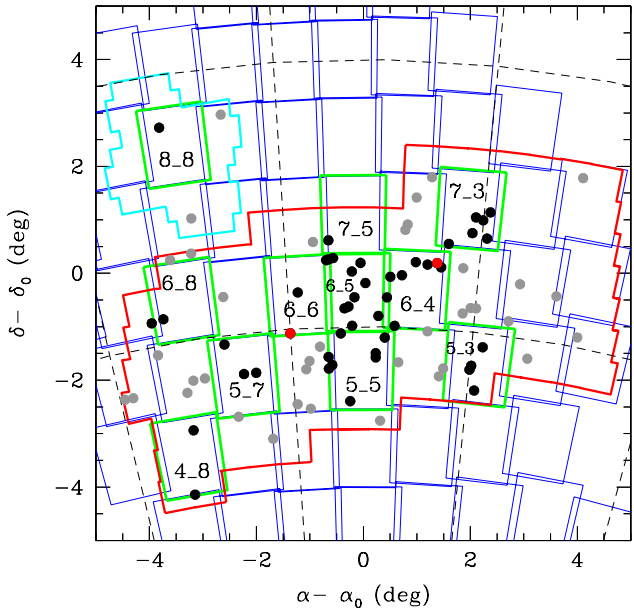}
\caption{Distribution of the known ACs over the LMC (projected in the
  sky adopting $\alpha_0 = 81.0$ deg and $\delta_0 = -69.0$ deg). Grey and black
  filled circles show the ACs observed by OGLE and those falling in
  the VMC tiles considered in this paper, respectively. The red filled
  circles represent the only two stars without a VMC counterpart within 1.0
  arcsec (see text). Thin blue and thick green
  squares (distorted by the projection into the sky) show part of the
  VMC tiles in the LMC and the 11 tiles treated in this paper,
  respectively. The thick red and light blue lines show the areas covered by OGLE
  III and IV (released up to now), respectively.}
\label{figMap}
\end{figure}

\section{Anomalous Cepheids in the VMC survey}

ACs in the LMC were identified and characterized in the
optical bands by \citet[][and references therein]{sos08} as part of
the OGLE~III project\footnote{http://ogle.astrouw.edu.pl}. For the LMC 
tile 8\_8 (including the South Ecliptic Pole, or SEP) which lies
outside the area imaged by OGLE~III, we used results of an early
release of stage IV of the OGLE survey \citep{sos12}. 
In these surveys, a total of 86 ACs were found (83 by OGLE~III and  
3 by OGLE~IV), of which 66 are F  and 20 are FO pulsators.    

In this paper we present results for the ACs included in
eleven``tiles'' (1.5 sq. deg.) completely or nearly completely
observed, processed and catalogued by the VMC survey as of March
2013, namely the tiles LMC 4\_8, 5\_3, 5\_5, 5\_7, 6\_4,  6\_5, 6\_6,
6\_8, 7\_3, 7\_5, and 8\_8 (see Fig.~\ref{figMap}). Tile LMC 6\_6 is centred on the well known  30 Dor star
forming region, tiles LMC 5\_5, 6\_4 and 6\_5 are placed on the bar of the LMC, 
whereas the remaining tiles lie in less crowded regions of the
galaxy.  We note
that tile LMC 8\_8 encompasses the South Ecliptic Pole (SEP) which will be
observed by the Gaia satellite during its commissioning phase (\citealt{Lin96,Lin10}).

The general observing strategy of the VMC survey is described in detail in Paper I,
whereas the procedures specifically applied to the variable stars can
be found in \citet{Moretti13}. Here we only recall that to obtain well 
sampled light curves,  the VMC $K_\mathrm{s}$-band time series observations were scheduled
in 12 separate epochs distributed over ideally several consecutive
months.  The VMC data, processed through the pipeline
\citep{Irwin_etal04} of the VISTA Data Flow System
\citep[VDFS,][]{Emerson_etal04} are in the VISTA photometric system 
(Vegamag=0).
The time-series photometry used in this paper 
was retrieved from the VISTA Science
Archive \citep[VSA,][]{Cross12}\footnote{http://horus.roe.ac.uk/vsa/}.
For our analysis we used the VMC data acquired until the end of March 2013.

According to OGLE~III/IV, 48 ACs are expected to lie in the 11 tiles
analysed in this paper. Figure~\ref{figMap} and Tab.~\ref{tabLog} show the distribution of
such stars in the VMC tiles. 
The OGLE~III and IV catalogues of ACs were cross-correlated against the VMC catalogue 
to obtain the $K_\mathrm{s}$ light curves for these variables.

\begin{figure*}
\includegraphics[height=15cm]{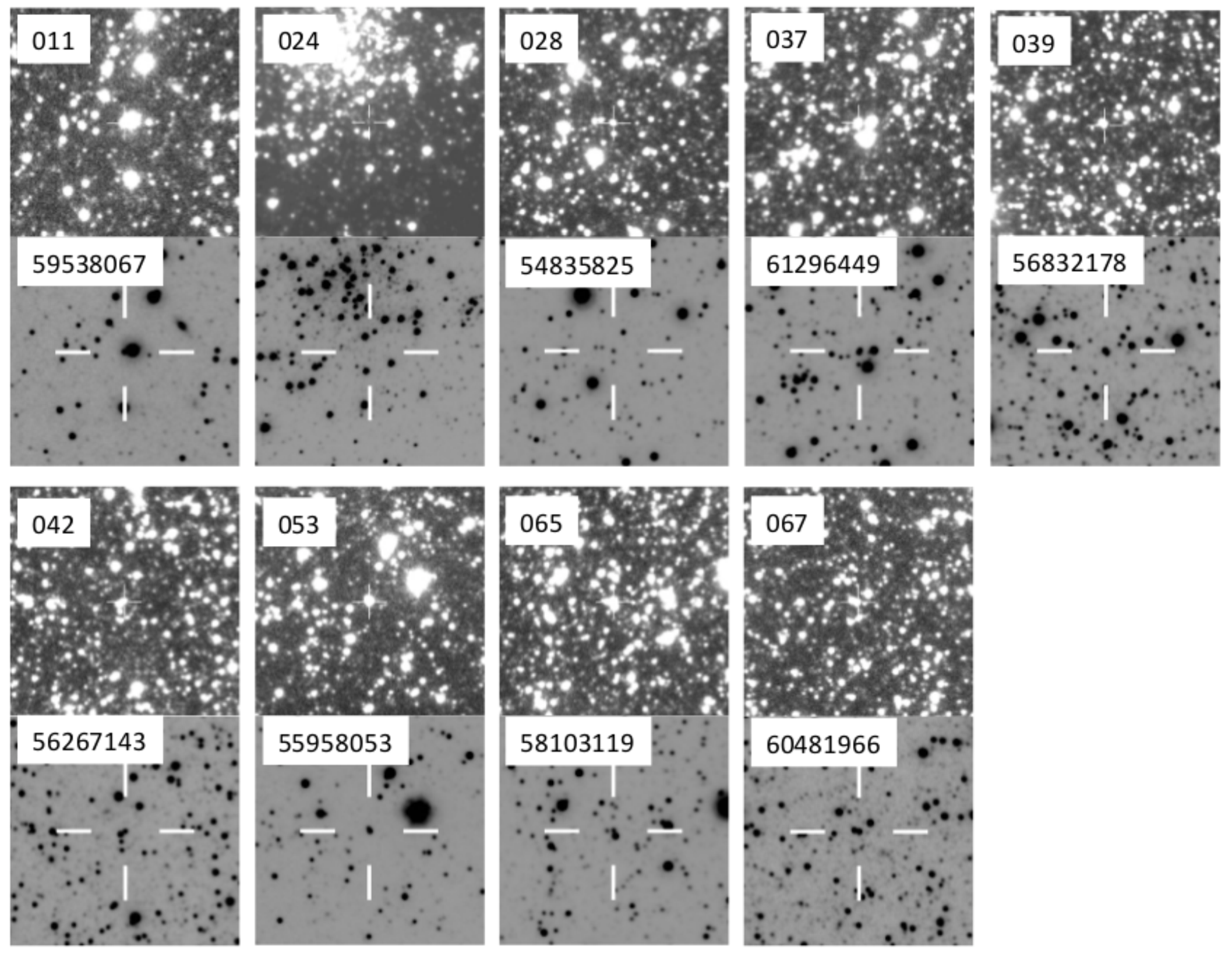}
\caption{Sky pictures for nine problematic stars extracted from the VMC (lower panels) 
 and the OGLE~III (upper panels) archives. The target is identified in the single-epoch VMC images by a
 label showing the last 8 digits of the VMC identification. Similarly the corresponding
 OGLE III identification is reported with three digits 
(i.e. without the prefix ``OGLE-LMC-AC-'').} 
\label{figCharts}
\end{figure*}

All but one of the 48 ACs were found to have a counterpart in the VMC
catalogue (see Tab.~\ref{tabData}). 
The only object without VMC counterpart in the VSA within 3 arcsec (see Fig.~\ref{figCharts}) 
is OGLE-LMC-ACEP-024, which is expected to fall 
well within the tile LMC 6\_4.
This star has has a luminosity $I$=17.679 mag and $V$=17.986 mag and no remarks in the OGLEIII catalogue. 
 There is a match if we enlarge the pairing radius
to 5 arcsec, but the corresponding star is clearly too bright
($K_\mathrm{s}\sim$12.8 mag). An inspection of the VMC image reveals that
the star is placed in the outskirts of the cluster NGC 1252, in a
rather crowded region. This could explain the lack of a detection. 
For one of the objects with VMC identification, i.e. OGLE-LMC-ACEP-065
in tile LMC 6\_6, the VSA database did not return any time-series
data. This is likely because the star is located at the very edge of the frame, where the
sensitivity is low and the target cannot be detected in the single
epoch frames. Looking in  more detail at this object on the VMC image (see
Fig.~\ref{figCharts}), it appears that some crowding is present but
the target should be easily detected at least in the neighbour tile
LMC 5\_6 which was not observed yet.  
\\

As shown in Tab.~\ref{tabData}, the sample of ACs discussed here includes 36 F and 10 FO-mode
pulsators. This sample represents more than  50\%
of the total number of ACs in the LMC. 
The VMC time-series $K_\mathrm{s}$ photometry for these 46 objects is provided
in Table~\ref{sampleTimeSeries}, which is published in its entirety in
the on-line version of the paper.

\begin{table}
\scriptsize
\caption{Number of Anomalous Cepheids in the 11 VMC tiles  analyzed in this paper, according to OGLE~III/IV.}
\label{tabLog}
\begin{center}
\begin{tabular}{cccccc}
\hline
\noalign{\smallskip} 
Tile & RA (center)  & DEC (center) & \#ACs & Epochs & OGLE \\
LMC       & J(2000)   &   J(2000) &  &(VMC) & \\
\noalign{\smallskip}
\hline
\noalign{\smallskip}  
4\_8   & 06:06:32.95  & $-$72:08:31.2 & 2  & 10 & III \\ 
5\_3   & 04:58:11.66 &  $-$70:35:28.0 & 4  & 11 & III \\ 
5\_5   & 05:24:30.34 &  $-$70:48:34.2 & 8  & 15 & III \\  
5\_7   & 05:51:04.87 &  $-$70:47:31.2 & 3  &  8  & III \\  
6\_4   & 05:12:55.80 &  $-$69:16:39.4 & 8  & 14 & III \\ 
6\_5   & 05:25:16.27 &  $-$69:21:08.3 & 10 & 9 & III \\ 
6\_6   & 05:37:40.01 &  $-$69:22:18.1 & 3  & 14 & III \\ 
6\_8   & 06:02:21.98 &  $-$69:14:42.4 & 2  & 14 & III \\ 
7\_3   & 05:02:55.20 &  $-$67:42:14.8 & 6  & 16 & III  \\ 
7\_5   & 05:25:58.44 &  $-$67:53:42.0 & 1  & 14 & III  \\ 
8\_8  &  05:59:23.14 &  $-$66:20:28.7 & 1  & 16 & IV  \\
\noalign{\smallskip}
\hline
\noalign{\smallskip}
\end{tabular}
\end{center}
\end{table}

\begin{table*}
\scriptsize\tiny
\caption{Cross-identification and main characteristics of the Anomalous Cepheids in the 11 ``tiles'' analysed in this paper.  
The columns report: 1) OGLE identification; 2) right ascension (OGLE); 3) declination (OGLE); 4) mode of pulsation; 5) period; 6) epoch of maximum light; 
7) intensity--averaged $I$ magnitude (OGLE); 8) intensity--averaged $V$ magnitude (OGLE); 9) VMC identification; 10) VMC Tile; 11) Number of $K_\mathrm{s}$ epochs; 12) Notes on individual stars}
\label{tabData}
\begin{center}
\begin{tabular}{cccccccccccc}
\hline
\noalign{\smallskip}   
ID & RA  & DEC  & M &  Period & Epoch &$\langle I \rangle$ & $\langle V \rangle$ & VMC-ID & Tile & n$_{\rm Epochs}$ & Notes\\
   & J2000  & J2000  & & d &  d &   mag & mag &  &  &  & \\                     
(1)    & (2)  & (3) & (4) & (5)&(6)  & (7) &(8)  &(9)  &  (10)&(11)  &(12) \\                     
\noalign{\smallskip}
\hline
\noalign{\smallskip}   
OGLE-LMC-ACEP-035  &  5:18:22.19  &  $-$69:03:38.4  &  FO  &  0.446049  &  52115.88331  &  18.143  &  18.658  &  558355019633  &  6\_4  &  14  &        \\
OGLE-LMC-ACEP-013  &  4:59:56.62  &  $-$70:42:24.9  &  FO  &  0.500923  &  52166.70918  &  17.963  &  18.488  &  558359106778  &  5\_3  &  11  &        \\
OGLE-LMC-ACEP-043  &  5:24:35.40  &  $-$68:48:22.6  &  FO  &  0.50647  &  52167.59135  &  17.913  &  18.593  &  558356009732  &  6\_5  &  9    &        \\
OGLE-LMC-ACEP-028  &  5:13:10.02  &  $-$68:46:11.9  &  FO  &  0.599253  &  50457.56427  &  17.574  &  18.322  &  558354835825  &  6\_4  &  14  & a       \\
OGLE-LMC-ACEP-068  &  5:51:09.02  &  $-$70:45:42.6  &  F  &  0.625645  &  52194.18981  &  18.265  &  18.809  &  558360489508  &  5\_7  &  8    &        \\
OGLE-LMC-ACEP-049  &  5:28:03.58  &  $-$69:39:15.2  &  F  &  0.644796  &  52167.32917  &  18.011  &  18.451  &  558356666394  &  6\_5  &  9    &        \\
OGLE-LMC-ACEP-071  &  5:54:43.24  &  $-$70:10:16.1  &  FO  &  0.676209  &  52187.35869  &  17.330  &  17.755  &  558360071488  &  5\_7  &  8   &        \\
OGLE-LMC-ACEP-045  &  5:26:24.68  &  $-$68:57:55.0  &  F  &  0.678431  &  52167.43289  &  18.325  &  18.976  &  558356105371  &  6\_5  &  9    &        \\
OGLE-LMC-ACEP-051  &  5:30:14.23  &  $-$68:42:31.1  &  F  &  0.708606  &  52167.2697  &  18.200  &  18.845  &  558355948009  &  6\_5  &  9     &        \\
OGLE-LMC-ACEP-023  &  5:07:52.42  &  $-$68:50:28.8  &  FO  &  0.723433  &  50726.2685  &  17.194  &  17.763  &  558354917807  &  6\_4  &  14   &        \\
OGLE-LMC-ACEP-034  &  5:17:11.15  &  $-$69:58:33.1  &  F  &  0.734298  &  52123.77557  &  17.874  &  18.474  &  558355708277  &  6\_4  &  14   &        \\
OGLE-LMC-ACEP-008  &  4:58:24.59  &  $-$71:05:11.5  &  FO  &  0.749068  &  52166.14121  &  17.300  &  17.911  &  558359341685  &  5\_3  &  11  &        \\
OGLE-LMC-ACEP-024  &  5:08:44.05  &  $-$68:46:01.2  &  F  &  0.794464  &  50726.76849  &  17.679  &  17.986  &  558354864345  &  6\_4  &  0  &     b \\
OGLE-LMC-ACEP-009  &  4:58:51.29  &  $-$67:44:23.9  &  FO  &  0.80008  &  52166.34228  &  17.344  &  17.913  &  558351428166  &  7\_3  &  16   &        \\
OGLE-LMC-ACEP-081  &  6:09:38.40  &  $-$69:34:04.1  &  F  &  0.800838  &  52194.39949  &  17.917  &  18.504  &  558354433783  &  6\_8  &  14   &        \\
OGLE-LMC-ACEP-067  &  5:48:22.08  &  $-$70:45:49.3  &  F  &  0.820921  &  52168.84504  &  17.786  &  18.451  &  558360481966  &  5\_7  &  7    &     c   \\
OGLE-LMC-ACEP-010  &  4:59:00.10  &  $-$68:14:01.1  &  F  &  0.834201  &  52166.34125  &  18.068  &  18.723  &  558351741077  &  7\_3  &  16   &        \\
OGLE-LMC-ACEP-078  &  6:06:58.20  &  $-$72:52:08.7  &  FO  &  0.856556  &  52187.11324  &  16.979  &  17.417  &  558367620049  &  4\_8  &  10  &        \\
OGLE-LMC-ACEP-041  &  5:21:14.35  &  $-$70:29:39.5  &  F  &  0.878142  &  50831.14818  &  17.625  &  18.200  &  558361532603  &  5\_5  &  15   &        \\
OGLE-LMC-ACEP-063  &  5:37:54.39  &  $-$69:19:28.7  &  F  &  0.893031  &  52187.44348  &  18.013  &  18.756  &  558357554473  &  6\_6  &  14   &        \\
OGLE-LMC-ACEP-007  &  4:57:31.48  &  $-$70:15:53.6  &  F  &  0.896399  &  52166.18452  &  17.688  &  18.247  &  558358821009  &  5\_3  &  11   &        \\
OGLE-LMC-ACEP-017  &  5:02:03.13  &  $-$68:09:30.4  &  F  &  0.929995  &  52166.07849  &  17.585  &  18.190  &  558351666932  &  7\_3  &  16   &        \\
OGLE-LMC-ACEP-040  &  5:21:13.22  &  $-$70:34:20.7  &  F  &  0.960577  &  50831.32386  &  17.433  &  18.037  &  558361590075  &  5\_5  &  15   &        \\
OGLE-LMC-ACEP-054  &  5:31:06.72  &  $-$68:22:29.8  &  F  &  0.980222  &  52167.47834  &  17.901  &  18.798  &  558353586650  &  7\_5  &  14   &        \\
OGLE-LMC-ACEP-039  &  5:20:44.47  &  $-$69:47:46.5  &  F  &  0.992407  &  50455.25698  &  17.660  &  18.245  &  558356832178  &  6\_5  &  9    &    c    \\
OGLE-LMC-ACEP-011  &  4:59:38.09  &  $-$70:37:45.5  &  F  &  0.99859  &  51999.6182  &  17.671  &  18.254  &  558359538067  &  5\_3  &  3      &      c  \\
OGLE-LMC-ACEP-050  &  5:28:57.71  &  $-$70:07:15.5  &  FO  &  1.044691  &  50454.74962  &  16.609  &  17.049  &  558361189093  &  5\_5  &  15  &        \\
OGLE-LMC-ACEP-042  &  5:23:34.60  &  $-$69:10:58.2  &  F  &  1.079036  &  52167.36992  &  17.897  &  18.715  &  558356267143  &  6\_5  &  5    &     c   \\
        LMC571.05.5070  &  6:01:41.77  &  $-$65:58:53.5  &  F  &  1.087061  &  55557.82471  &  17.181  &  17.686  &  558349409852  &  8\_8  &  16      &        \\
OGLE-LMC-ACEP-077  &  6:04:35.73  &  $-$71:40:35.8  &  F  &  1.122498  &  52187.66068  &  17.459  &  18.099  &  558367137280  &  4\_8  &  10   &        \\
OGLE-LMC-ACEP-056  &  5:31:49.45  &  $-$70:33:22.6  &  F  &  1.124003  &  52167.63676  &  17.284  &  17.877  &  558361558150  &  5\_5  &  15   &        \\
OGLE-LMC-ACEP-079  &  6:07:02.01  &  $-$69:31:55.2  &  F  &  1.15517  &  52176.92756  &  17.149  &  17.634  &  558354405855  &  6\_8  &  14    &        \\
OGLE-LMC-ACEP-037  &  5:19:16.67  &  $-$70:11:58.4  &  F  &  1.25774  &  52156.5207  &  17.132  &  17.743  &  558361296449  &  5\_5  &  15     &     a   \\
OGLE-LMC-ACEP-036  &  5:18:58.85  &  $-$69:26:47.8  &  F  &  1.257982  &  50455.67396  &  17.160  &  17.787  &  558355324744  &  6\_4  &  23   &        \\
OGLE-LMC-ACEP-052  &  5:31:01.53  &  $-$70:42:22.2  &  F  &  1.262555  &  52167.0715  &  17.008  &  17.577  &  558361664456  &  5\_5  &  15    &        \\
OGLE-LMC-ACEP-046  &  5:26:27.17  &  $-$69:58:57.0  &  F  &  1.263717  &  50454.39955  &  17.264  &  17.851  &  558356990050  &  6\_5  &  9    &        \\
OGLE-LMC-ACEP-021  &  5:06:37.49  &  $-$68:23:40.3  &  F  &  1.295843  &  50725.62518  &  17.188  &  17.827  &  558351789022  &  7\_3  &  16   &        \\
OGLE-LMC-ACEP-044  &  5:25:54.11  &  $-$69:26:52.9  &  F  &  1.308509  &  50455.10639  &  17.052  &  17.609  &  558356474792  &  6\_5  &  9    &        \\
OGLE-LMC-ACEP-032  &  5:15:56.13  &  $-$69:01:29.1  &  F  &  1.316022  &  50456.81167  &  17.167  &  17.780  &  558355007611  &  6\_4  &  14   &        \\
OGLE-LMC-ACEP-065  &  5:40:03.04  &  $-$70:04:47.8  &  F  &  1.3215432  &  50725.78028  &  17.041  &  17.508  &  558358103119  &  6\_6  &  0   &   b     \\
OGLE-LMC-ACEP-016  &  5:01:36.69  &  $-$67:51:33.1  &  F  &  1.54567  &  52166.4502  &  16.926  &  17.448  &  558351480147  &  7\_3  &  16     &        \\
OGLE-LMC-ACEP-048  &  5:27:12.12  &  $-$69:37:19.6  &  F  &  1.545893  &  50454.48193  &  16.718  &  17.324  &  558356636279  &  6\_5  &  9    &        \\
OGLE-LMC-ACEP-055  &  5:31:41.11  &  $-$68:44:37.7  &  F  &  1.606665  &  52189.7241  &  17.011  &  17.603  &  558357237330  &  6\_6  &  14    &        \\
OGLE-LMC-ACEP-057  &  5:31:49.88  &  $-$70:46:30.0  &  F  &  1.710008  &  52167.36219  &  16.813  &  17.455  &  558361710363  &  5\_5  &  15   &        \\
OGLE-LMC-ACEP-026  &  5:10:42.62  &  $-$68:48:19.6  &  F  &  1.738745  &  50457.22581  &  16.816  &  17.483  &  558354874727  &  6\_4  &  14   &        \\
OGLE-LMC-ACEP-053  &  5:31:06.20  &  $-$68:43:45.3  &  F  &  1.888099  &  52166.63019  &  16.738  &  17.299  &  558355958053  &  6\_5  &  23   &   a     \\
OGLE-LMC-ACEP-047  &  5:27:05.27  &  $-$71:23:33.4  &  F  &  2.177985  &  52166.33026  &  16.881  &  17.482  &  558362082677  &  5\_5  &  15   &        \\
OGLE-LMC-ACEP-014  &  5:00:08.26  &  $-$67:54:04.3  &  F  &  2.291346  &  52164.63844  &  16.639  &  17.241  &  558351518318  &  7\_3  &  16   &        \\
\noalign{\smallskip}
\hline
\noalign{\smallskip}
\multicolumn{12}{l}{$^{\rm (a)}$ stars showing significant blending but with useful light curves (see Figs. ~\ref{figCharts}, ~\ref{figLC}, and Fig.~\ref{figLC1})}\\
\multicolumn{12}{l}{$^{\rm (b)}$ stars without $K_\mathrm{s}$ VMC data (see Fig.~\ref{figCharts})} \\
\multicolumn{12}{l}{$^{\rm (c)}$ stars showing significant blending and having unusable light curves (see Fig. ~\ref{figCharts} and ~\ref{figLC2})}  \\
\end{tabular}
\end{center}
\end{table*}

\begin{table}
\scriptsize
\caption{$K_\mathrm{s}$ time-series photometry of the ACs}
\label{sampleTimeSeries}
\begin{center}
\begin{tabular}{ccc}
\hline
\noalign{\smallskip} 
HJD-2\,400\,000 & $K_\mathrm{s}$  & $\sigma_{K_\mathrm{s}}$  \\
\noalign{\smallskip}
\hline
\noalign{\smallskip}
\multicolumn{3}{c}{AC OGLE-LMC-ACEP-007} \\
\noalign{\smallskip}
\hline
\noalign{\smallskip} 
 56267.81039 &    17.118 &     0.039   \\
 56316.64476 &    16.921 &     0.032   \\
 56318.55354 &    16.899 &     0.030   \\
 56322.63275 &    17.170 &     0.036   \\
 56328.57027 &    16.925 &     0.030   \\
 56334.54372 &    16.933 &     0.036   \\
 56341.53132 &    17.200 &     0.042   \\
 56347.56046 &    17.036 &     0.032   \\
 56371.53252 &    16.923 &     0.034   \\
 56372.51981 &    16.952 &     0.034   \\
 56375.52238 &    17.112 &     0.034   \\
\noalign{\smallskip}
\hline
\noalign{\smallskip}
\end{tabular}
\end{center}
Table~\ref{sampleTimeSeries} is published in its entirety only in the
electronic edition of the journal. 
A portion is shown here for guidance regarding its form and content.
\end{table}

%

\begin{figure*}
\includegraphics[width=16cm]{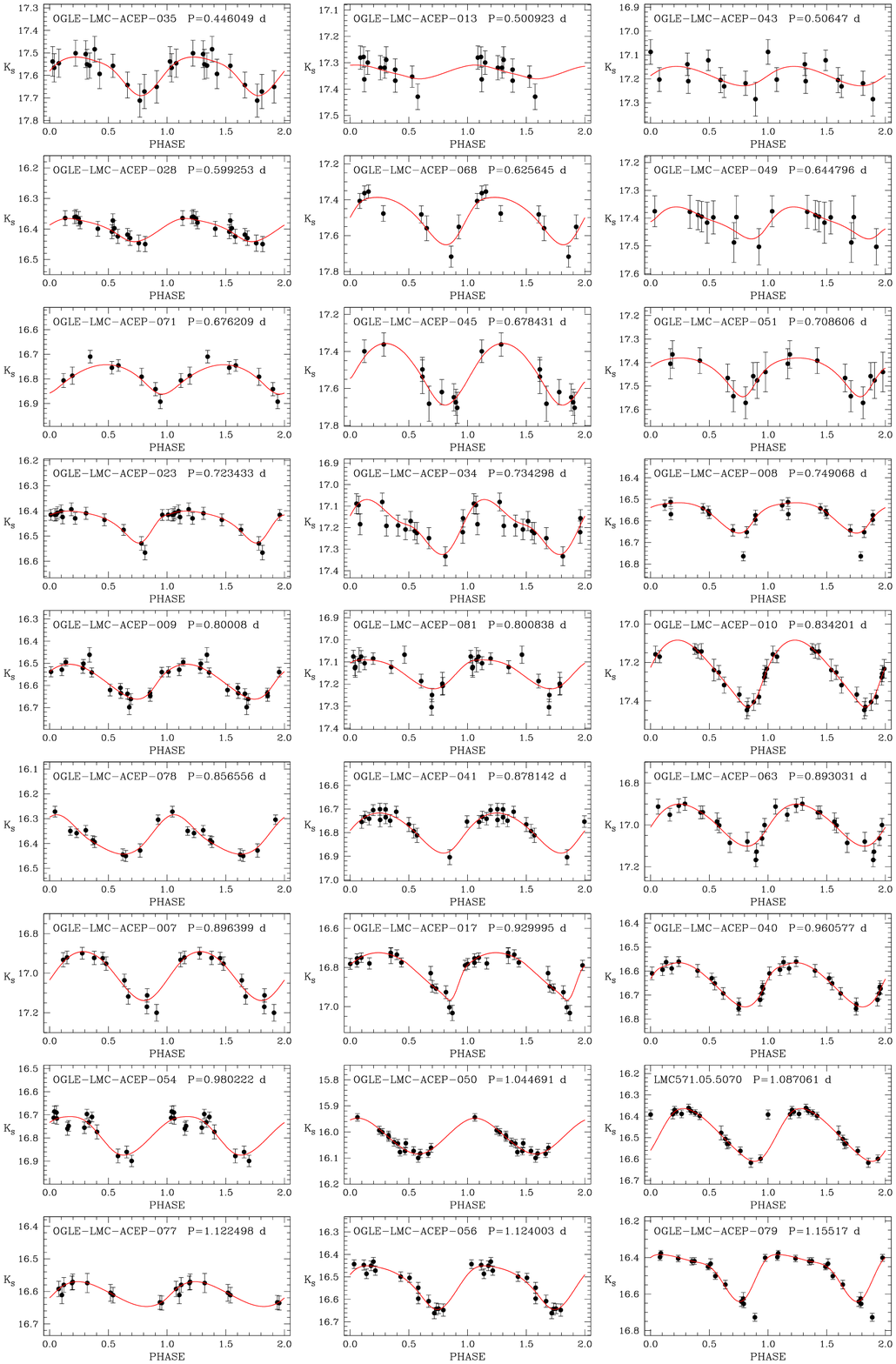}
\caption{$K_\mathrm{s}$--band light curves for 27 of the 42 ACs with
  usable data discussed in this paper. Stars are displayed in order of increasing
period. Solid lines represent spline best-fits to the data (see
text). In each panel we report OGLE's identification number  and 
period.} 
\label{figLC}
\end{figure*}

\begin{figure*}
\includegraphics[width=16cm]{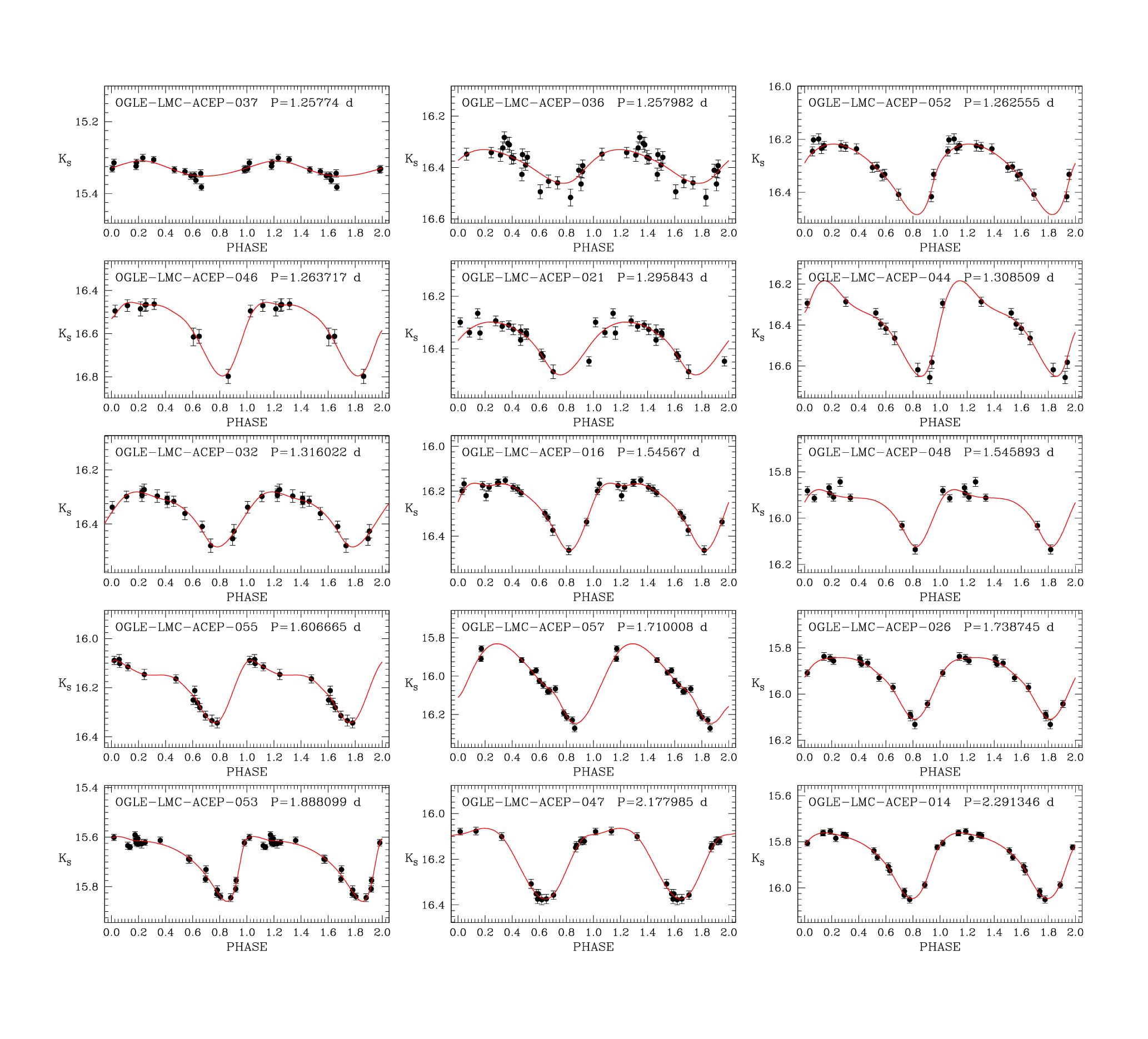}
\caption{As in Fig.~\ref{figLC} but for the remaining 15 ACs of our
sample.} 
\label{figLC1}
\end{figure*}

Periods and Epochs of maximum light available from the 
OGLE~III catalogue were used to  fold the
$K_\mathrm{s}$--band light curves produced by the VMC
observations. The OGLE~IV catalogue provides only the period for the AC LMC571.05.5070,  hence we obtained
the Epoch of maximum from the analysis of the star $V$-band light curve.
The $K_\mathrm{s}$--band light curves for a sample of 42 ACs with
useful light curves are shown in Figs.~\ref{figLC} and \ref{figLC1}. 
Apart from a few cases these light curves are generally well sampled and nicely shaped. 
Intensity-averaged $\langle K_\mathrm{s} \rangle $ magnitudes were derived from the light curves
simply using custom software written in {\sc C}, that performs
a spline interpolation to the data with no need of using templates. 
Some evidently discrepant data points in
the light curves were excluded from the fit but were plotted in the 
figure for completeness (note that most of these ``bad'' data points
belong to  observations collected during nights that did not strictly
meet the VMC quality criteria). The final spline fit to the data is shown
by a solid line in Figs. ~\ref{figLC} and \ref{figLC1}.
Final $\langle K_\mathrm{s} 
\rangle$ magnitudes are provided in Table~\ref{tabResults}.  

\begin{figure}
\includegraphics[width=8.5cm]{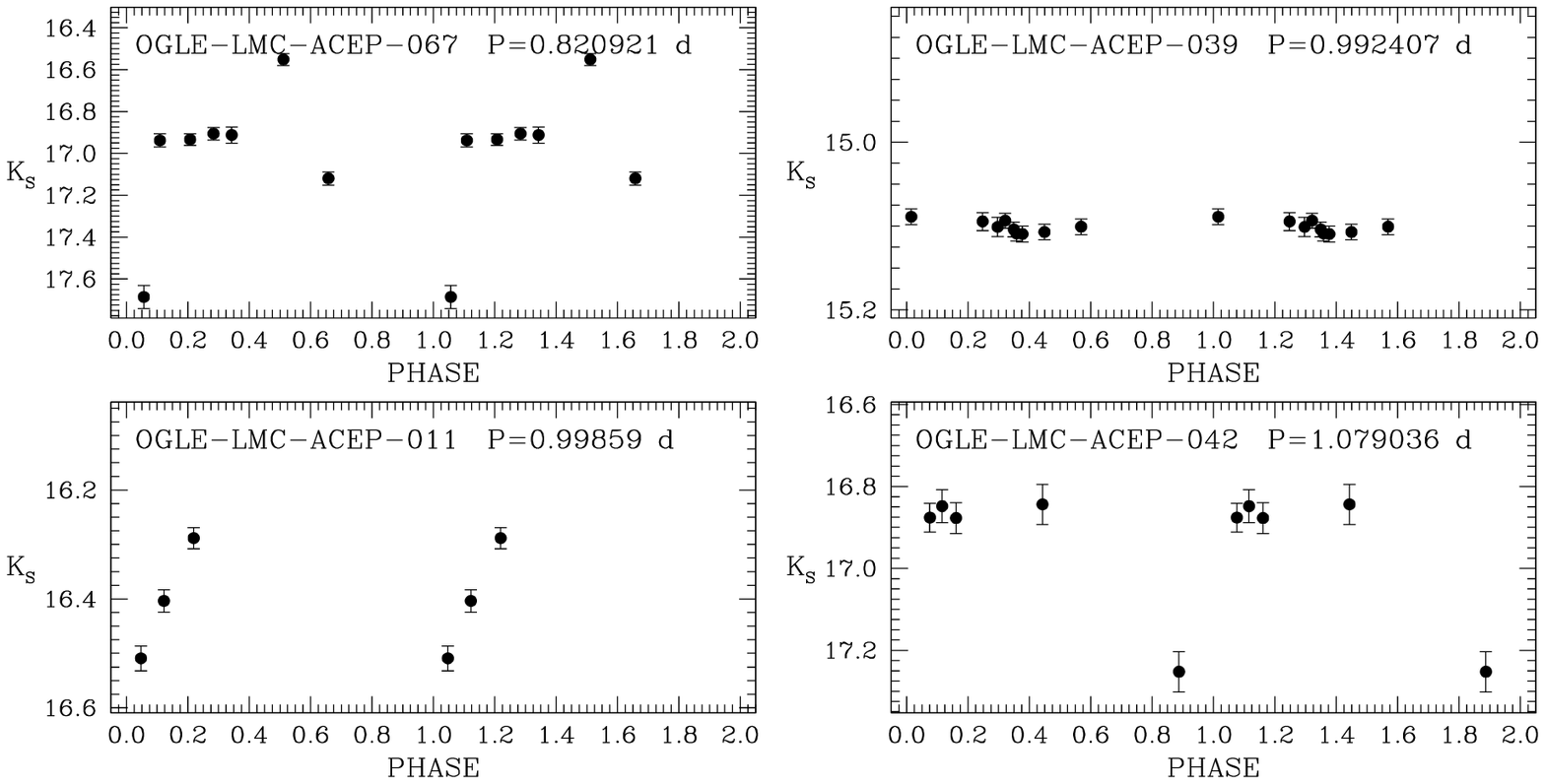}
\caption{Light curves for four problematic stars (see text).} 
\label{figLC2}
\end{figure}

Four objects in our sample were excluded, namely: OGLE-LMC-ACEP-011,
039, 042, 067. 
Their   light curves are displayed  in Fig.~\ref{figLC2}, whereas their finding
charts are shown in Fig.~\ref{figCharts}. A quick analysis of the finding charts 
reveals that all these stars have significant problems of
crowding/blending (particularly,  star 011).

\begin{table}
\scriptsize\tiny
\caption{Results for the 42 Anomalous Cepheids with useful light curves analysed in this paper.  
The columns report: 1) OGLE identification; 2) mode of pulsation; 3) period; 4) intensity--averaged $K_\mathrm{s}$ magnitude; 5) uncertainty on the $\langle K_\mathrm{s} \rangle$;  
6) peak--to--peak amplitude in $ K_\mathrm{s}$ ; 7) adopted reddening; 8) correction in magnitude due to the deprojection.}
\label{tabResults}
\begin{center}
\begin{tabular}{c@{ }c@{ }c@{ }c@{ }c@{ }c@{ }c@{ }c@{ }}
\hline
\noalign{\smallskip}   
ID & M & Period & $\langle K_\mathrm{s} \rangle$ & $\sigma_{\langle K_\mathrm{s} \rangle}$ &  $Amp(K_\mathrm{s})$ & $E($V$$-$$I$)$ & $\Delta$ \\
   & & d & mag & mag &mag & mag &mag\\                     
(1)    & (2)  & (3) & (4) & (5) &(6)  & (7) &(8)  \\                     
\noalign{\smallskip}
\hline
\noalign{\smallskip}   
OGLE-LMC-ACEP-035  &  FO  &  0.446049  &  17.585  &  0.026  &  0.17  &  0.07  &  0.026                   \\
OGLE-LMC-ACEP-013  &  FO  &  0.500923  &  17.334  &  0.033  &  0.06  &  0.03  &  0.068                  \\
OGLE-LMC-ACEP-043  &  FO  &  0.50647  &  17.186  &  0.051  &  0.08  &  0.09  &  0.013                   \\
OGLE-LMC-ACEP-028  &  FO  &  0.599253  &  16.399  &  0.010  &  0.08  &  0.06  &  0.040                   \\
OGLE-LMC-ACEP-068  &  F  &  0.625645  &  17.490  &  0.049  &  0.26  &  0.06  &  $-$0.042                  \\
OGLE-LMC-ACEP-049  &  F  &  0.644796  &  17.412  &  0.028  &  0.12  &  0.04  &  0.003                 \\
OGLE-LMC-ACEP-071  &  FO  &  0.676209  &  16.790  &  0.015  &  0.12  &  0.07  &  $-$0.053                 \\
OGLE-LMC-ACEP-045  &  F  &  0.678431  &  17.499  &  0.045  &  0.33  &  0.11  &  0.008                 \\
OGLE-LMC-ACEP-051  &  F  &  0.708606  &  17.437  &  0.023  &  0.17  &  0.10  &  0.000                     \\
OGLE-LMC-ACEP-023  &  FO  &  0.723433  &  16.444  &  0.005  &  0.13  &  0.08  &  0.052                  \\
OGLE-LMC-ACEP-034  &  F  &  0.734298  &  17.186  &  0.022  &  0.26  &  0.13  &  0.028                   \\
OGLE-LMC-ACEP-008  &  FO  &  0.749068  &  16.567  &  0.007  &  0.14  &  0.08  &  0.072                  \\
OGLE-LMC-ACEP-009  &  FO  &  0.80008  &  16.575  &  0.014  &  0.16  &  0.06  &  0.081                   \\
OGLE-LMC-ACEP-081  &  F  &  0.800838  &  17.147  &  0.017  &  0.13  &  0.07  &  $-$0.090                   \\
OGLE-LMC-ACEP-010  &  F  &  0.834201  &  17.225  &  0.015  &  0.35  &  0.11  &  0.077                   \\
OGLE-LMC-ACEP-078  &  FO  &  0.856556  &  16.372  &  0.018  &  0.16  &  0.09  &  $-$0.057                 \\
OGLE-LMC-ACEP-041  &  F  &  0.878142  &  16.790  &  0.020  &  0.17  &  0.09  &  0.021                   \\
OGLE-LMC-ACEP-063  &  F  &  0.893031  &  16.990  &  0.026  &  0.20  &  0.18  &  $-$0.020                    \\
OGLE-LMC-ACEP-007  &  F  &  0.896399  &  17.002  &  0.020  &  0.25  &  0.06  &  0.073                   \\
OGLE-LMC-ACEP-017  &  F  &  0.929995  &  16.810  &  0.015  &  0.25  &  0.06  &  0.070                    \\
OGLE-LMC-ACEP-040  &  F  &  0.960577  &  16.644  &  0.012  &  0.18  &  0.09  &  0.021                   \\
OGLE-LMC-ACEP-054  &  F  &  0.980222  &  16.781  &  0.029  &  0.17  &  0.26  &  $-$0.000                     \\
OGLE-LMC-ACEP-050  &  FO  &  1.044691  &  16.016  &  0.011  &  0.14  &  0.07  &  0.003                 \\
LMC571.05.5070  &  F  &  1.087061  &  16.479  &  0.007  &  0.25  &  0.08  &  $-$0.072                     \\
OGLE-LMC-ACEP-077  &  F  &  1.122498  &  16.608  &  0.002  &  0.08  &  0.07  &  $-$0.062                  \\
OGLE-LMC-ACEP-056  &  F  &  1.124003  &  16.527  &  0.019  &  0.19  &  0.06  &  $-$0.002                 \\
OGLE-LMC-ACEP-079  &  F  &  1.15517  &  16.471  &  0.013  &  0.26  &  0.05  &  $-$0.085                   \\
OGLE-LMC-ACEP-037  &  F  &  1.25774  &  15.332  &  0.007  &  0.04  &  0.08  &  0.024                    \\
OGLE-LMC-ACEP-036  &  F  &  1.257982  &  16.388  &  0.022  &  0.13  &  0.08  &  0.023                   \\
OGLE-LMC-ACEP-052  &  F  &  1.262555  &  16.316  &  0.011  &  0.27  &  0.07  &  0.001                  \\
OGLE-LMC-ACEP-046  &  F  &  1.263717  &  16.570  &  0.016  &  0.35  &  0.03  &  0.008                  \\
OGLE-LMC-ACEP-021  &  F  &  1.295843  &  16.377  &  0.027  &  0.20  &  0.07  &  0.058                   \\
OGLE-LMC-ACEP-044  &  F  &  1.308509  &  16.380  &  0.024  &  0.48  &  0.06  &  0.007                  \\
OGLE-LMC-ACEP-032  &  F  &  1.316022  &  16.360  &  0.010  &  0.20  &  0.06  &  0.032                   \\
OGLE-LMC-ACEP-016  &  F  &  1.54567  &  16.261  &  0.006  &  0.30  &  0.05  &  0.073                    \\
OGLE-LMC-ACEP-048  &  F  &  1.545893  &  15.958  &  0.018  &  0.25  &  0.07  &  0.005                  \\
OGLE-LMC-ACEP-055  &  F  &  1.606665  &  16.188  &  0.010  &  0.26  &  0.08  &  $-$0.003                 \\
OGLE-LMC-ACEP-057  &  F  &  1.710008  &  16.006  &  0.017  &  0.43  &  0.07  &  $-$0.001                 \\
OGLE-LMC-ACEP-026  &  F  &  1.738745  &  15.932  &  0.011  &  0.27  &  0.08  &  0.046                   \\
OGLE-LMC-ACEP-053  &  F  &  1.888099  &  15.682  &  0.011  &  0.26  &  0.11  &  $-$0.002                 \\
OGLE-LMC-ACEP-047  &  F  &  2.177985  &  16.181  &  0.009  &  0.31  &  0.07  &  0.012                   \\
OGLE-LMC-ACEP-014  &  F  &  2.291346  &  15.862  &  0.008  &  0.29  &  0.04  &  0.076                   \\
\noalign{\smallskip}
\hline
\noalign{\smallskip}
\end{tabular}
\end{center}
\end{table}


It is important to remember that all the $K_\mathrm{s}$ photometry
presented in this paper is in the VISTA system. To make it easy to compare
our results with the widely used 2MASS system, we note that the two
systems are very close to each other. In particular, the VMC
$K_\mathrm{s}$ magnitude depends only mildly on 
the ($J$$-$$K_\mathrm{s}$) colour. Indeed, the empirical results available
to date\footnote{http://casu.ast.cam.ac.uk/surveys-projects/vista/technical/photometric-properties}  
show that: ($J$$-$$K_\mathrm{s}$)(2MASS)=1.081($J$-$K_\mathrm{s}$)(VISTA)
and $K_\mathrm{s}$(2MASS)=$K_\mathrm{s}$(VISTA)$-$0.011($J$-$K_\mathrm{s}$)(VISTA). 
Unfortunately, for the majority of our targets we only have a few $J$ measurements
(4 phase points), hence a star-by-star correction based on the  $(\langle
J \rangle-\langle K_\mathrm{s} \rangle)$ colour is likely to introduce larger errors than the correction itself.  
Furthermore, since the measured ($\langle$$J$ $\rangle$$-$$\langle$ $K_\mathrm{s}$$
\rangle$) of our AC sample typically ranges from 0.2 to 0.5 mag, the average
correction over the 42 ACs considered here, is as small as $-3.0\pm$1.0 mmag and can be safely neglected.  
In conclusion, for ACs, as well as for CCs \citep[see][]{Ripepi12b}, to a very good approximation, the VISTA and 2MASS
$K_\mathrm{s}$ can be considered equivalent.

\section{Optical Period-Luminosity, Period-Luminosity-Colour and Period-Wesenheit relations}

Before presenting the results in the NIR bands, in this section we derive 
 the coefficients of the optical ($V,I$) $P$--$L$, $P$--$L$--$C$ and
Wesenheit--$W(V,I)$ relations, based on the OGLE data, since they are
still lacking in the literature. Indeed, neither \citet{sos08}
nor \citet{fiomon12} published these relations, although they showed
them in some figures. In this case we have used the whole sample of 
ACs detected in the LMC by the OGLE III/IV surveys. 
 
The first step is to correct for  reddening, which unfortunately is rather
variable in the LMC, hence needs to be evaluated locally. To this aim, we adopted the recent estimates by
\citet{haschke}. Individual  $E$($V$$-$$I$) reddening values  for the
42 ACs with useful VMC data are reported in column 7 of Table~\ref{tabResults}. 
In Sect. 5 we will check the soundness of these reddening values.

The second step consists of accounting for the inclination of the LMC disc-like structure
 by de-projecting each AC with respect to the LMC centre. We followed the procedure 
outlined in \citet{marel01} and adopted their values 
of the LMC centre, inclination, and position angle of the line of
nodes (see column 8 of Tab.~\ref{tabResults}).  

\begin{figure*}
\includegraphics[width=16cm]{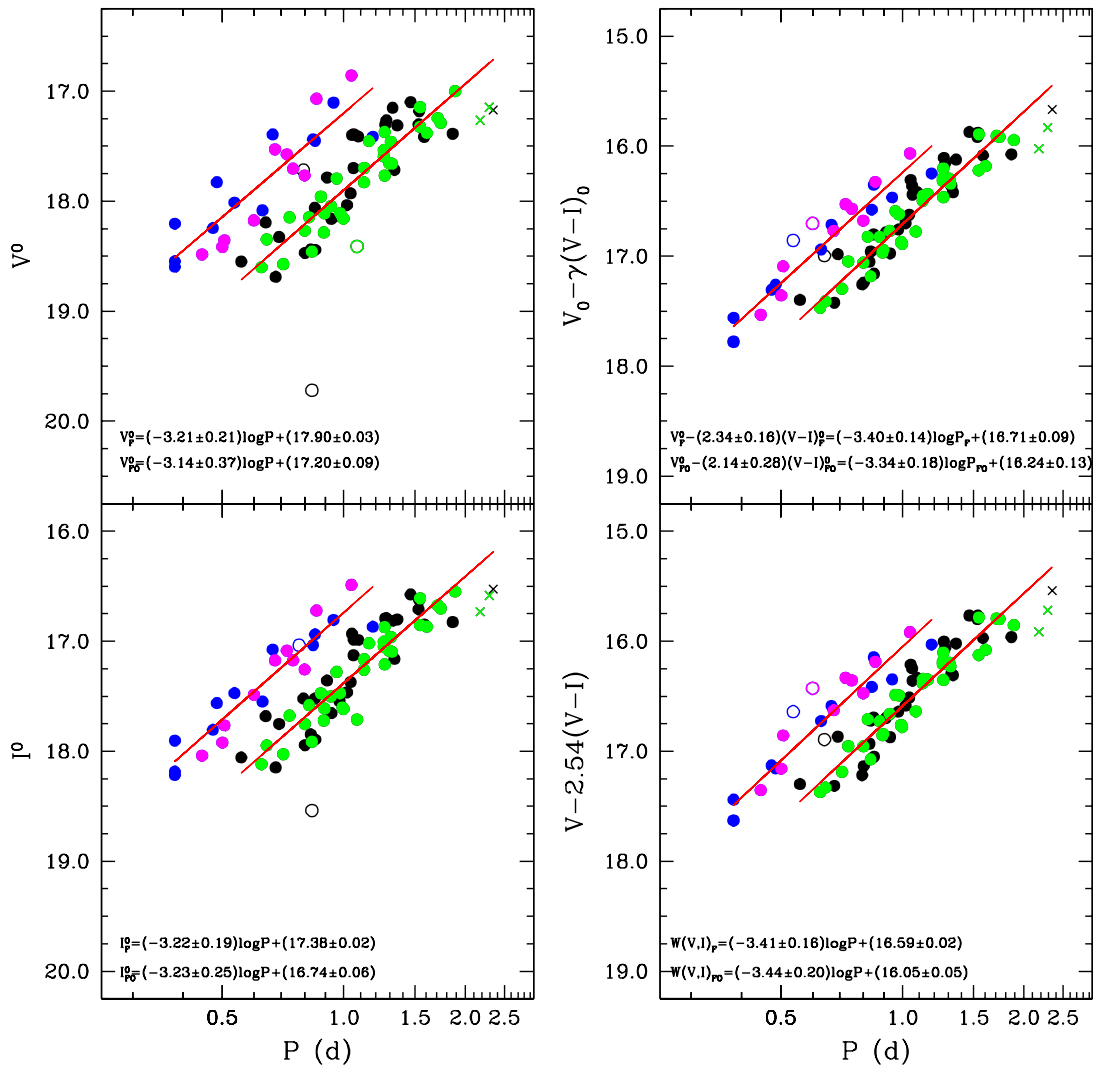}
\caption{Optical $P$--$L$, $P$--$W$ and $P$--$L$--$C$ relations for F and FO Anomalous
 Cepheids.  The Wesenheit function is defined as:
$W(V, I)=V-2.54(V-I)$. In each panel,  green and black  filled circles are OGLE  F
pulsators, blue and magenta filled circles are OGLE FO pulsators,  with (in green and magenta) or  without (in black and blue) VMC $K_\mathrm{s}$ measurements. 
Empty circles and crosses mark pulsators that were discarded in the
derivation of the $P$--$L$ and $P$--$W$ (and the optical $P$--$L$--$C$) relations. In particular, the three stars with longest periods (crosses) 
likely are not ACs and rather  belong to a different type of variability. 
In all panels, the  red lines show the best-fits to the
data, with the corresponding relationships  labelled at the
bottom.} 
\label{figOptical}
\end{figure*}

Finally, we performed least-squares fits to the data of  F and
FO-mode variables separately, adopting equations of the form 
Mag=$\alpha+\beta$ log$P$, with Mag$=V,I$. 
The best-fitting relationships are shown in the left panels of Fig.~\ref{figOptical}, their 
coefficients are provided in the first portion of Table~\ref{tabOptical}.
We note that in all panels of this figure, the stars plotted with  crosses and open circles  were 
rejected from the fit because they are 2.5-3 $\sigma$ off 
the regression line. In particular, the three objects  with the  longest
periods, namely OGLE-LMC-ACEP-014, 033, 047 (crosses in
Fig.~\ref{figOptical})   are likely not ACs 
but rather  members of some different variable classes such as BL
Herculis stars. Indeed, looking at Fig. 1 in \citet{sos08} it is clear
that some confusion can be possible between ACs and BL Her pulsators
at the periods of interest.  
Additional stars that were not used in at least one of the regressions are
OGLE-LMC-ACEP-022, 024, 028, 042, 059, 083. All these objects, with the
possible exception of 022, show some observational problems. In particular,
the stars 024, 042, 028 will be discussed in detail in the next section,
since they are problematic also in the VMC data (see
Fig.~\ref{figCharts}). As for the remaining three, we inspected the
OGLE images and light curves, finding that in the case of the star 083
there is a strong background variation close to the object (possibly
close to an edge of the CCD) and the star light curves appear to be rather noisy. Stars 022
and 059 are surrounded by a number of close companions and both of them 
show a rather noisy light curve, especially star 059, that has also a
large $(V-I)$ colour ($>$1.2 mag).

In addition to  the  $V, I$ $P$--$L$ relations  we can also derive the $P$--$W$ and 
$P$--$L$--$C$ relations. The advantages of  using these relations instead of
a simple $P$--$L$ have been widely discussed in the
literature \citep[see e.g.][in the case of ACs]{mfc04}. 
Briefly, these relations include a colour term with a coefficient that, in the
case of the $P$--$L$--$C$ relation, takes into account the colour distribution
of the variable stars within the instability strip, whereas in the
case of the Wesenheit function  corresponds to the ratio between total-to-selective extinction 
in the filter pair \citep{m82,Caputo00}, thus
making the Wesenheit relations reddening free by definition. We expect these 
relations  to have much smaller dispersion than a simple $P$--$L$ relation,
even if the scatter reduction for ACs is not as significant as in the case of
CCs \citep[see e.g.][]{mmf05,mfc04}. 
In fact,  for CCs a strict Mass-Luminosity ($ML$) relation is predicted to exist 
by  stellar evolution computations for He burning
intermediate mass stars, which makes the $P$--$L$--$C$ a relation holding for
each individual pulsator \citep[see e.g.][for details]{Caputo00}.
Unfortunately, the ACs are not characterized by such a strict $ML$ relation, thus 
the resulting $P$--$L$--$C$ and $P$--$W$ relations include the possible effect of mass
differences at fixed luminosity level. \\
The $P$--$W$ and $P$--$L$--$C$ relations are usually calculated using the $(V-I)$
colour.  The coefficients of the  relations
derived with this procedure for the LMC ACs are provided in the lower
portions of Table~\ref{tabOptical}. The relations are shown in
the right  panels of Fig.~\ref{figOptical}.
The dispersion of the
$P$--$W$ and $P$--$L$--$C$ relations is of the order of $\sim$ 0.15 mag (see Tab.~\ref{tabOptical}), hence 
smaller than for the $P$--$L$ relation but larger than in the
case of CCs \citep[see e.g.][who
found a $\sigma$=0.08 mag for the $P$--$W(V,I)$ relation]{sos08}.

\begin{table}
\small
\caption{Optical $P$--$L$, $P$--$W$ and $P$--$L$--$C$ relations for F and FO Anomalous
Cepheids.  The Wesenheit function is defined as:
$W(V, I)=V-2.54(V-I)$.}
\label{tabOptical}
\begin{center}
\begin{tabular}{cccccccc}
\hline
\noalign{\smallskip}
mode & $\alpha$ & $\sigma_{\alpha}$ & $\beta$ & $\sigma_{\beta}$ & $\gamma$ & $\sigma_{\gamma}$ &r.m.s. \\
\noalign{\smallskip}
\hline
\noalign{\smallskip}
\multicolumn{8}{c}{$V^0$=$\alpha$+$\beta$ log$P$} \\
\noalign{\smallskip}
\hline
\noalign{\smallskip}
F   & 17.90 & 0.03 & $-$3.21 & 0.21 & & & 0.20  \\
FO & 17.20 & 0.09 & $-$3.14 & 0.37 & &  & 0.23\\
\noalign{\smallskip}
\hline
\noalign{\smallskip}
\multicolumn{8}{c}{$I^0$=$\alpha$+$\beta$ log$P$} \\
\noalign{\smallskip}
\hline
\noalign{\smallskip}
F   & 17.38 & 0.02 & $-$3.22 & 0.19 & & & 0.18  \\
FO & 16.74 & 0.06 & $-$3.23 & 0.25 & &  & 0.16\\
\noalign{\smallskip}
\hline
\noalign{\smallskip}
\multicolumn{8}{c}{$W(V,I)=\alpha+\beta $ log$P$} \\
\noalign{\smallskip}
\hline
\noalign{\smallskip}
F & 16.59 & 0.02 & $-$3.41 & 0.16& &  & 0.15\\
FO   & 16.05 & 0.05 & $-$3.44 & 0.22 & & & 0.13  \\
\noalign{\smallskip}
\hline
\noalign{\smallskip}
\multicolumn{8}{c}{$V^0=\alpha+\beta$ log$P$$+\gamma ($V$$-$$I$)^0$} \\
\noalign{\smallskip}
\hline
\noalign{\smallskip}
F   & 16.71 & 0.09 & $-$3.40 & 0.14 & 2.34& 0.16& 0.14  \\
FO & 16.24 & 0.13 & $-$3.34 & 0.18 & 2.14& 0.28 & 0.12\\
\noalign{\smallskip}
\hline
\noalign{\smallskip}
\end{tabular}
\end{center}
\end{table}

\begin{table}
\small
\caption{NIR $P$--$L$ and  $P$--$W$ relations for F and FO-mode Anomalous
Cepheids.  The Wesenheit function is defined as:
$W(V, K)$=$K_\mathrm{s}$$-$0.13 ($V$$-$$K_\mathrm{s}$). Note that all the results are in
the VISTA photometric system.}
\label{NIR}
\begin{center}
\begin{tabular}{cccccccc}
\hline
\noalign{\smallskip}
mode & $\alpha$ & $\sigma_{\alpha}$ & $\beta$ & $\sigma_{\beta}$ & $\gamma$ & $\sigma_{\gamma}$ &r.m.s. \\
\noalign{\smallskip}
\hline
\noalign{\smallskip}
\multicolumn{8}{c}{$K_\mathrm{s}^0$=$\alpha$+$\beta$ log$P$} \\
\noalign{\smallskip}
\hline
\noalign{\smallskip}
F & 16.74 & 0.02 & $-$3.54 & 0.15& &  & 0.10\\
FO & 16.06 & 0.07 & $-$4.18 & 0.33& &  & 0.10\\
\noalign{\smallskip}
\hline
\noalign{\smallskip}\multicolumn{8}{c}{$W(V,K_\mathrm{s})=\alpha+\beta $ log$P$} \\
\noalign{\smallskip}
\hline
\noalign{\smallskip}
F & 16.58 & 0.02 & $-$3.58 & 0.15& &  & 0.10\\
F & 15.93 & 0.07 & $-$4.14 & 0.33& &  & 0.10\\
\noalign{\smallskip}
\hline
\noalign{\smallskip}
\end{tabular}
\end{center}
\end{table}

\section{$K_\mathrm{s}$-band Period-Luminosity and Period-Wesenheit relations}

\begin{figure}
\includegraphics[width=8.8cm]{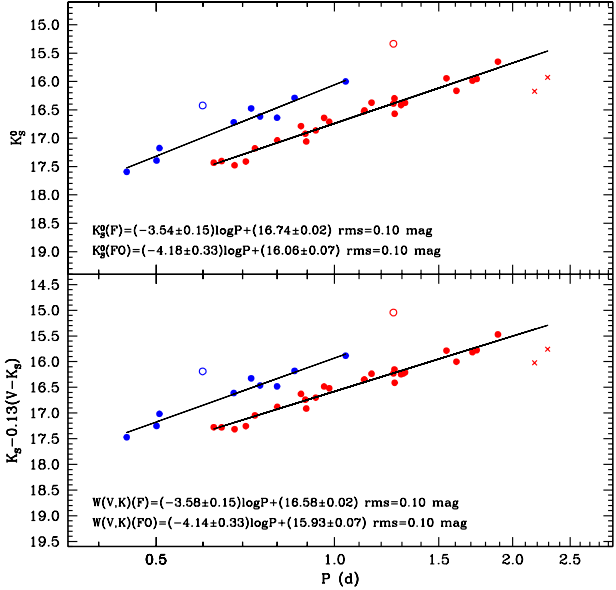}
\caption{Top panel: $K_\mathrm{s}$-band $P$--$L$ relation for F and
  FO-mode ACs (red and blue symbols, respectively). 
Filled circles show the objects used for the least 
square fits. Stars marked by open circles and crosses were 
discarded (see text for detail). The solid lines is the least 
squares fit to the data. Bottom panel: as for the top panel but for the 
Wesenheit function, which  is defined as labelled in the figure.} 
\label{figNIR}
\end{figure}

By analogy to the optical $V,I$ bands, we can calculate the $P$--$L$,
and $P$--$W$ relationships in the $K_\mathrm{s}$-band. Our sample
consist of 32 F-mode and 10 FO-mode ACs that are sufficient to provide statistically
meaningful $P$--$L$ and $P$--$W$ relations. For both modes of
pulsation our sample maps the whole range of periods
covered by the LMC ACs, as shown in Fig.~\ref{figOptical},
where we have plotted in green and magenta colours the F and FO-mode
objects observed by VMC, respectively. 
The $K_\mathrm{s}$-band $P$--$L$ and $P$--$W$ relations obtained  are displayed in Fig.~\ref{figNIR}, whereas 
their coefficients are summarized in Tab.~\ref{NIR}.  Three F-mode and
one FO-mode stars were excluded from the fit 
because they are more than 2$\sigma$ off the regression lines. The
stars OGLE-LMC-ACEP-037 and 028, a F and FO-mode pulsators
respectively, are shown with empty circles in Fig.~\ref{figNIR}. They
have rather ``normal'' light curves, but at least the star 037 is
clearly blended  (see Fig.~\ref{figCharts}). The star 028 also exhibit an unusual 
colour (see Sect. 5 and Fig.~\ref{figMass}) and we 
hypothesize that it may be blended, however,  we do not have enough 
resolution to detect the contaminant star. 
The two additional F-Mode ACs excluded from the fit (namely, OGLE-LMC-ACEP-014 and OGLE-LMC-ACEP-047, shown as crosses
in Fig.~\ref{figNIR}) are the stars with
the longest periods. They  were  excluded also in the optical analysis as we suspect they could possibly belong to a separate class of variables. 
We note that, as for CCs, moving to the NIR   
the dispersion of  the AC $P$--$L$ and $W(V,K_\mathrm{s})$ relations 
decreases.
We also made an attempt  to calculate $P$--$L$--$C$ relations in the
form $K_\mathrm{s}^0$=$f({\rm log}P,
(V-K_\mathrm{s})_0)$, but the colour term coefficient turned out to be statistically
equal to zero. This is not very surprising given the 
relatively small number of pulsators  and the  almost identical dispersions of 
the $P$--$L$ and $P$--$W$ relations. Hence, according to this dataset,
in the NIR filters it seems that 
there is no great advantage in using the AC $P$--$L$--$C$ or pseudo $P$--$L$--$C$
relations instead of the $P$--$L$.

\section{Comparison with literature and application to real cases}

\begin{figure}
\includegraphics[width=8.8cm]{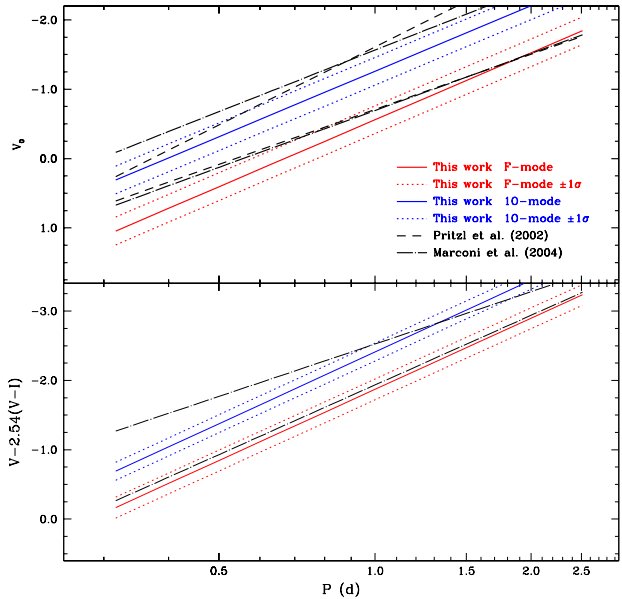}
\caption{Comparison between the optical $P$--$L$  (top panel) and
  $P$--$W(V,I)$ relations (bottom panel)
for F and FO-mode ACs derived in this paper with the literature results.} 
\label{compLiterature}
\end{figure}

To compare our 
$P$--$L$s for ACs with the relations  available in the literature, we 
first need to set the absolute zero points. This can be done by assuming 
a proper value for the distance to the host galaxy, the LMC.
We have adopted our own evaluation based on the absolute calibrations of the $P$--$L$, $P$--$L$--$C$ and
$W(V,K_\mathrm{s})$ relations for CCs  in the LMC presented in 
\citet{Ripepi12b}. In that paper we used the trigonometric parallaxes of  
Galactic Cepheids as well as Baade-Wesselink measurements of Cepheids in
the LMC to evaluate $DM_{\rm LMC}^0=18.46\pm0.03$ mag \citep[see][for
full details.]{Ripepi12b}. 
This distance modulus  is in very good
agreement with current literature estimates for the distance to the LMC
as nicely reviewed by \citet{w12}, and with the very recent 
and precise value by  \citet{Pietrzynski2013}, based on eclipsing binaries. 
Results of this comparison in the optical are summarized in the first part of Tab.~\ref{comparison} and graphically shown in
Fig.~\ref{compLiterature} for the $P$--$L$ and the Wesenheit relations, respectively.

The upper panel of Fig.~\ref{compLiterature} shows the comparison 
of our $P$--$L$s with the relations derived empirically by \citet{Pritzl02} 
and \citet{mfc04} on the basis of a few tens of ACs belonging to a number of 
dwarf Spheroidal galaxies (dSph) orbiting the Milky Way and M31 spirals. There
is a clear disagreement between our results and the literature. This
is likely due to a number of reasons: i) the very different coverage of the AC instability strip,
which is much more uniformly and completely covered in the LMC sample; ii)  at shorter periods
(P$\sim$0.4-0.5 d) it is easy to confuse F-- and FO--mode  pulsators,
especially in  the dSphs where samples are generally rather small;
iii) the non-homogeneous dSph sample.
A better agreement, at least for the F-mode pulsators, can be seen in
the bottom panel of   Fig.~\ref{compLiterature}, where we compare our
$W(V,I)$  function with  those by \citet{mfc04}, the only published
relation to
date. The better agreement in this case is probably due to the much lower
dependence of the $W(V,I)$ with respect to the $P$--$L$ on the way the
pulsators populate the instability strip. 

\citet{mfc04} also published mass-dependent,
$P$--$L$--$C$ and Wesenheit (the latter only for F-mode pulsators) relations 
for ACs calculated on the basis of
non--linear, non--local time-dependent convective pulsation models.  
The lower portion of Tab.~\ref{comparison} reports the comparison between \citet{mfc04} 
theoretical models and our results.  
There is good agreement for the slope of the  $P$--$W(V,I)$ relations, however, our zero
point implies a stellar mass of $M\sim1.3 M_{\odot}$, at the lower end of the allowed range for
ACs. 

As for the $P$--$L$--$C$, both the slopes and the zero point disagree 
by more than 1 $\sigma$. We do not have a clear explanation for these discrepancies, 
however,  if the reddening values we have  adopted for the ACs are correct, the disagreement 
could in principle be related to uncertainties in the theoretical colour-temperature
relations.

\begin{figure}
\includegraphics[width=8.8cm]{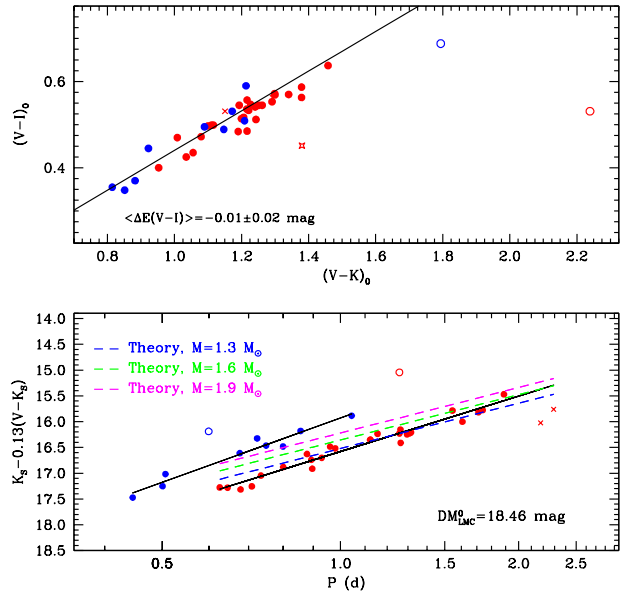}
\caption{Top panel: $(V-I)_0$ versus $(V-K_\mathrm{s})_0$ plot for the F-mode 
ACs analysed in this paper. Symbols are as in
Fig.~\ref{figNIR}. The solid line represents the theoretical relation
by \citet{mfc04}.  The theoretical $(V-K_\mathrm{s})_0$ was
converted to the 2MASS system using the relations in Tab.~\ref{comparison}.
Bottom panel: comparison between the observed Wesenheit
$K_\mathrm{s}$-band relation for F-mode (solid line; note that the
FO-mode relation is shown, too)  and the predictions by \citet{mfc04}  for three different choices of 
 the pulsator mass (dashed lines, see labels).} 
\label{figMass}
\end{figure}

There are no other empirical NIR relations in the literature we are aware of, 
 hence we can  compare 
our results only with the colour-colour and
the $P$--$W(V, K_\mathrm{s})$ relations derived by \citet{mfc04} from the 
 theory of stellar pulsation. 
 The top panel of
Fig.~\ref{figMass} shows this comparison for the  42 pulsators analysed in this paper in
the $(V-I)_0$$-$$(V-K_\mathrm{s})_0$ plane. 
Overall the agreement is very satisfactory. The
residual discrepancy can be minimized  by a difference  
in the adopted reddening of only $\langle \Delta E(V-I) \rangle=-0.01\pm0.02$
mag. This is a confirmation that the reddening values adopted in
this paper are well--established. An inspection of the figure 
confirms that the two stars OGLE-LMC-ACEP-037 and 028 (empty
circles), previously excluded from the $P$--$L$ and $P$--$W$
derivation, are highly deviant also in the colour-colour plane. In
addition, we find a third discrepant star in this plane: 
OGLE-LMC-ACEP-053 (empty star). Also this object is clearly blended
(see Fig.~\ref{figCharts}), so that its strange position in the
colour-colour plane can be easily explained. Since this AC shows a
rather normal light curve and it is not deviant e.g. in the $P$--$L$
relation, we conclude that its photometry  is only mildly affected by 
the blending star.

The lower panel of Fig.~\ref{figMass} shows the comparison between our
F-mode $P$--$W(V, K_\mathrm{s})$ relation and theoretical predictions
(not available for FO-mode pulsators), for three different
choices of the AC mass encompassing approximately the whole range
of allowed values for this parameter, namely $M=1.3, 1.6$ and $1.9
M_{\odot}$. There is a good agreement for the slope of the
$P$--$W(V, K_\mathrm{s})$ relations, whereas, as already found for the optical  $P$--$W(V,I)$,  our zero-point seems to
favor the smallest value of $M=1.3 M_{\odot}$, for the mass.

\begin{table*}
\small
\caption{Literature values for the coefficients of the $P$--$L$, $P$--$W$ and $P$--$L$--$C$ relations, for F and FO Anomalous 
Cepheids.  The Wesenheit functions are 
defined as: $W(V, I)=V-2.54(V-I)$ and 
$W(V,K_\mathrm{s})=K_\mathrm{s}-0.13(V-K_\mathrm{s})$. In the
$K_\mathrm{s}$, the photometry of previous
studies was converted to the 2MASS system, for consistency with
our results (see Section 2).}
\label{comparison}
\begin{center}
\begin{tabular}{ccccccccl}
\hline
\noalign{\smallskip}
mode & $\alpha$ & $\sigma_{\alpha}$ & $\beta$ & $\sigma_{\beta}$ & $\gamma$ & $\sigma_{\gamma}$ &r.m.s. & source \\
\noalign{\smallskip}
\hline
\noalign{\smallskip}
\multicolumn{9}{c}{$M_V^0$=$\alpha$+$\beta$ log$P$} \\
\noalign{\smallskip}
\hline
\noalign{\smallskip}
F   &  $-0.56$ & 0.04 & $-3.21$& 0.21&  & & 0.20& This paper  \\ 
F   & $-0.71$ & 0.03   & $-2.64$   & 0.17  &  &  &   & \citet{Pritzl02} \\
F   & $-0.70$ & 0.14 &$-2.73$&  &  & & 0.14&\citet{mfc04}$^{\rm (a)}$\\   
FO & $-1.26$& 0.09 & $-3.14$& 0.37&  & & 0.23& This paper \\
FO & $-1.61$& 0.07 & $-3.74$& 0.20 &  & & 0.25&\citet{Pritzl02}  \\
FO & $-1.58$& 0.25 & $-2.95$&  &  & & 0.25&\citet{mfc04}$^{\rm (a)}$  \\
\noalign{\smallskip}
\hline
\noalign{\smallskip}
\multicolumn{9}{c}{$W(V,I)=\alpha+\beta $ log$P$} \\
\noalign{\smallskip}
\hline
\noalign{\smallskip}
F   & $-1.87$ & 0.04 & $-3.41$& 0.16&  & & 0.15& This paper  \\  
F   & $-1.93$ & 0.20 & $-3.34$& &  & & 0.20&\citet{mfc04}$^{\rm (a)}$  \\  
F   & $-1.74-1.83$$\log{M}$ & 0.20 & $-3.34$& &  & & 0.20&\citet{mfc04}$^{\rm (b)}$  \\  
FO   & $-2.41$ & 0.06 & $-3.44$& 0.22&  & & 0.13& This paper  \\  
FO  & $-2.52$ & 0.20 & $-2.51$&  &  & & 0.20&\citet{mfc04}$^{\rm (a)}$  \\  
\noalign{\smallskip}
\hline
\noalign{\smallskip}
\multicolumn{9}{c}{$M_V^0=\alpha+\beta$ log$P+\gamma (V-I)_0$} \\
\noalign{\smallskip}
\hline
\noalign{\smallskip}
F   & $-1.75$ & 0.09 & $-3.40$ & 0.14 & 2.34& 0.16& 0.14  & This paper \\
F   & $-1.83-1.89$$\log {M}$ &  & $-2.93$ &  & 2.73 & & 0.01  &\citet{mfc04}$^{\rm (b)}$ \\
FO & $-2.22$ & 0.13 & $-3.34$ & 0.18 & 2.14 & 0.28 & 0.12 & This paper\\
FO   & $-2.24-1.73$$\log {M}$  &  & $-3.09$ &  & 2.67 & & 0.01  &\citet{mfc04}$^{\rm (b)}$ \\
\noalign{\smallskip}
\hline
\hline
\noalign{\smallskip}
\multicolumn{9}{c}{$K_\mathrm{s}=\alpha+\beta $ log$P$} \\
\noalign{\smallskip}
\hline
\noalign{\smallskip}
F   & $-1.72$ & 0.04 & $-3.54$& 0.15&  & & 0.10& This paper  \\  
FO   & $-2.40$ & 0.07 & $-4.18$& 0.33&  & & 0.10& This paper  \\  
\noalign{\smallskip}
\hline
\noalign{\smallskip}
\multicolumn{9}{c}{$W(V,K_\mathrm{s})=\alpha+\beta $ log$P$} \\
\noalign{\smallskip}
\hline
\noalign{\smallskip}
F   & $-1.88$ & 0.04 & $-3.58$& 0.15&  & & 0.10&This paper  \\  
F   & $-1.71-1.83$$\log {M}$  &  & $-2.93$ &  &  & & 0.04  &\citet{mfc04}$^{\rm (b,c)}$ \\
FO   & $-2.53$ & 0.07 & $-4.14$& 0.33&  & & 0.10&This paper  \\  
\noalign{\smallskip}
\hline
\noalign{\smallskip}
\multicolumn{9}{l}{$^{\rm (a)}$ Empirical }\\
\multicolumn{9}{l}{$^{\rm (b)}$ Theoretical, the value of  the mass ranges
 from 1.3 to 1.9 $M_{\odot}$}\\
\multicolumn{9}{l}{$^{\rm (c)}$ Models transformed to the Johnson system:
 for ACs $K$(Johnson) $\approx K$(SAAO)} \\
\multicolumn{9}{l}{\citep[see][]{bessellbrett88}. $K_\mathrm{s}$(2MASS)$=K$(SAAO)$+0.02(J-K)$(SAAO)$-0.025$ \citep[][]{carpenter}} \\
\end{tabular}
\end{center}
\end{table*}

\subsection{Application of the optical $P$--$W$ relation}

In the previous sections we have shown that the $P-W(V,I)$ relation 
 is likely the best tool we have to use the ACs  as standard
candles. Indeed, the $P-W(V,I)$ relation does not depend on the reddening and on how
the pulsators populate the instability strip. This obviously holds also for
the $P-W(V,K_\mathrm{s})$. However, at present the lack of
NIR observations of ACs in other galaxies limits severely the use of this relation.
On the contrary,
$V,I$ data are available for a significant number of ACs belonging to a few dwarf 
galaxies in the Local Group that can be used to verify the ability of
our $P-W(V,I)$ to estimate the distance to the host systems by
comparing AC-based and RR Lyrae-based distance determinations. 

\begin{figure}
\includegraphics[width=8.8cm]{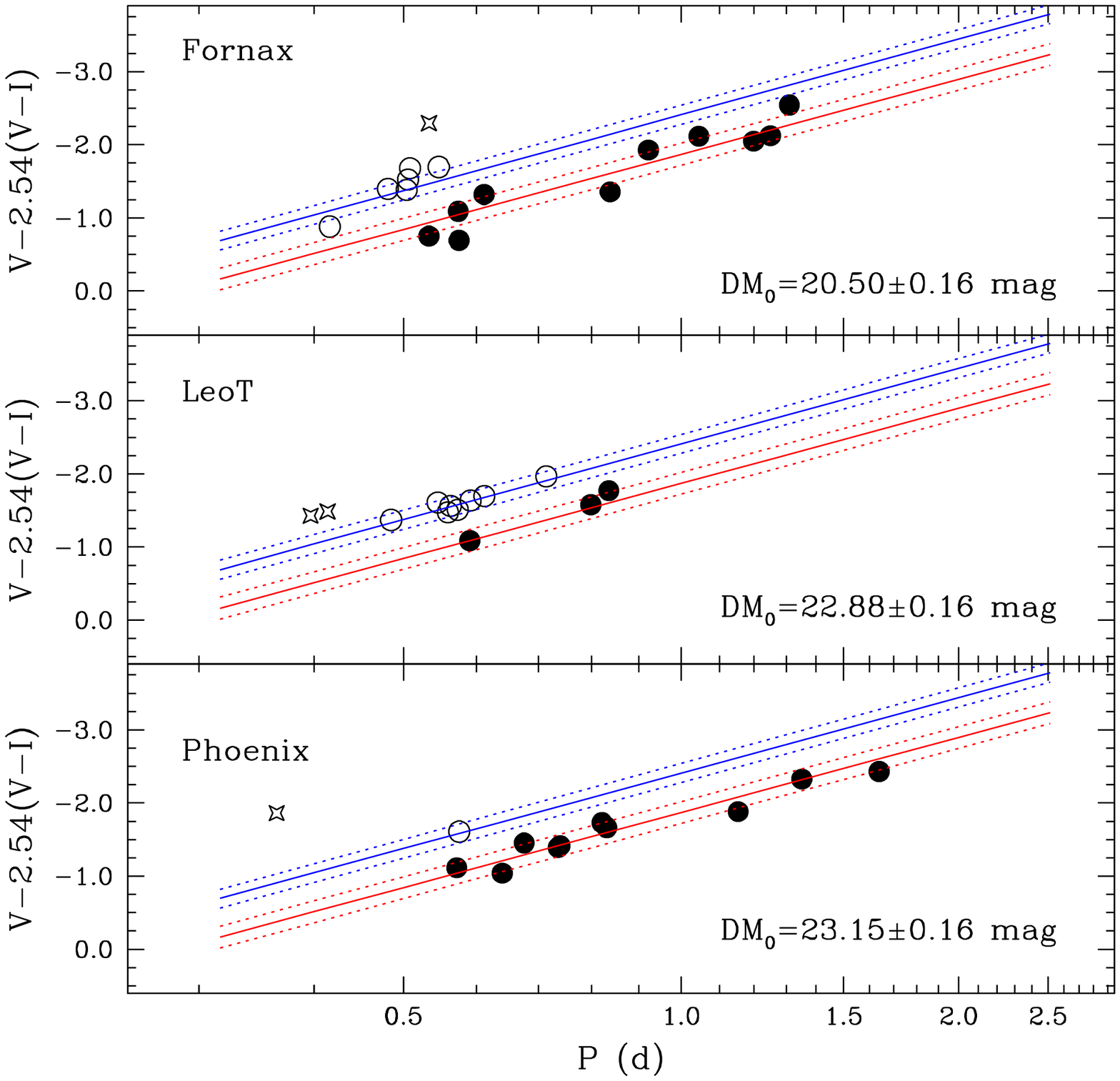}
\caption{Optical $P-W(V,I)$ relations for ACs in a number of Local
  Group dwarf galaxies in comparison with the results obtained in this
  paper.  
F and FO-mode ACs are marked by filled and empty circles, respectively;  starred symbols refer
to objects with uncertain classification that were not used in
calculating the distances.  Red and blue lines show present work $P-W(V,I)$ 
for F and FO-mode ACs, respectively.
True distance moduli estimated from these relations are labelled.} 
\label{plDsph}
\end{figure}

To our knowledge, there are only three dwarf galaxies with a significant number
of ACs measured in the $V,I$ bands, namely, Fornax \citep[][]{Bersier02}, 
Leo~T \citep{Clementini2012}, and Phoenix \citep{Gallart04}. There is no
clear separation between F and FO-mode pulsators in these galaxies
\citep[see][]{Pritzl02,mfc04}, hence we used our own $P$--$W$
relation to perform  a tentative subdivision of the samples into the two modes. The
result of the overall procedure is shown in Fig.~\ref{plDsph}, where
the different panels report the results for the three afore-mentioned galaxies. 
The distance moduli labelled in the figures were
obtained trying to adjust simultaneously F-mode and FO-mode
pulsators, and weighting the results to take into account 
the number of pulsators in each of the two modes. Errors on the 
derived distances are dominated by the dispersion of the LMC 
$P-W(V,I)$, whereas the uncertainty of the LMC distance is significantly smaller. 

In Table~\ref{compDistance} we compare the distances obtained from the ACs and 
literature values based on RR Lyrae or other distance indicators. 
Our distance modulus for the Fornax dSph is
systematically smaller than all the other determinations,
but  still in agreement with most of them within 1$\sigma$. 
On the contrary, the agreement is rather good for Leo~T and excellent  for Phoenix. 
On this basis, it is not easy to explain why we find such shorter
distances for Fornax dSph. New observations,
including the complete sample of ACs belonging to this very large
galaxy are needed to clarify this point. 

Concluding, the $P-W(V,I)$ derived in this paper is a reliable tool 
for the determination of the distance to galaxies hosting 
significant samples of ACs. 

\begin{table}
\small
\caption{Comparison between distances derived from the ACs in this
  paper and  the literature values for the Fornax, Leo~T, and Phoenix 
  Local Group dwarf.}
\label{compDistance}
\begin{center}
\begin{tabular}{llc}
\hline
\noalign{\smallskip}
method & $DM_0$  & Reference \\
\noalign{\smallskip}
\hline
\noalign{\smallskip}
\multicolumn{3}{c}{Fornax: $DM_0$(ACs)=20.50$\pm$0.16 mag (15 ACs)} \\
\noalign{\smallskip}
\hline
\noalign{\smallskip}
TRGB           & 20.76$\pm$0.20  & (1)  \\
Red Clump    & 20.66                  & (2) \\
 TRGB            & 20.65$\pm$0.11  & (2) \\
 Red Clump    & 20.86$\pm$0.01 & (3)\\
 RR Lyrae       & 20.72$\pm$0.10  & (4) \\
 TRGB            & 20.75$\pm$0.19  & (5) \\
 HB                & 20.70$\pm$0.12  & (6) \\ 
TRGB            & 20.84$\pm$0.18  & (7)  \\  
RR Lyrae       & 20.66$\pm$0.07   & (8)   \\ 
\noalign{\smallskip}
\hline
\noalign{\smallskip}
\multicolumn{3}{c}{Leo~T: $DM_0$(ACs)=22.88$\pm$0.16 mag (11 ACs)} \\
\noalign{\smallskip}
\hline
\noalign{\smallskip}
TRGB            & 23.1$\pm$0.2         & (9)  \\
SFH-Fitting  & 23.05                     & (10) \\
RR Lyrae       & 23.06$\pm$0.15   & (11)   \\ 
\noalign{\smallskip}
\hline
\noalign{\smallskip}
\multicolumn{3}{c}{Phoenix: $DM_0$(ACs)=23.15$\pm$0.16 mag (11 ACs)} \\
\noalign{\smallskip}
\hline
\noalign{\smallskip}
TRGB            & 23.0$\pm$0.1  & (12)   \\ 
TRGB            & 23.21$\pm$0.08  & (13)   \\ 
TRGB            & 23.11  & (14)   \\ 
Mira            & 23.10$\pm$0.18  & (15)   \\ 
TRGB            & 23.09$\pm$0.10  & (16)   \\ 
\noalign{\smallskip}
\hline
\multicolumn{3}{l}{\scriptsize (1) \citet{Buonanno1999}; (2) \citet{Bersier2000};} \\
\multicolumn{3}{l}{\scriptsize (3) \citet{Pietrzynski2003}; (4) \citet{Greco2005}; } \\
\multicolumn{3}{l}{\scriptsize (5) \citet{Gullieuszik2007}; (6) \citet{Rizzi2006}; } \\
\multicolumn{3}{l}{\scriptsize (7) \citet{Pietrzynski2009}; (8) \citet{Greco2009}; } \\
\multicolumn{3}{l}{\scriptsize (9) \citet{Irwin2007}; (10) \citet{Weisz2012}; } \\
\multicolumn{3}{l}{\scriptsize (11) \citet{Clementini2012};} \\
\multicolumn{3}{l}{\scriptsize (12) \citet{Martinez-Delgado1999}; } \\
\multicolumn{3}{l}{\scriptsize (13) \citet{Held1999}; } \\
\multicolumn{3}{l}{\scriptsize (14) \citet{Holtzman2000}; } \\
\multicolumn{3}{l}{\scriptsize (15) \citet{Menzies2008}; (16) \citet{Hidalgo2009}; } \\
\end{tabular}
\end{center}
\end{table}

\section{Summary and Conclusions}

We have presented the first light curves in the NIR $K_\mathrm{s}$-band
 for Anomalous Cepheids. In particular, our sample consists  of 46
AC pulsators (36 F-mode and 10 FO-mode) located in the LMC and observed 
by the VMC survey. 
Our light curves are well sampled with the number of epochs ranging 
from 8 to 23. In spite of the faintness of the LMC  ACs, these data 
allowed us to obtain very precise mean $K_\mathrm{s}$ magnitudes for 
the pulsators, with average errors of the order of 0.01 mag. 

The $\langle K_\mathrm{s} \rangle$ magnitudes were used to build the first $P$--$L$ and
$P$--$W$ relations in the NIR  for F and FO-mode ACs. At the same time we exploited the OGLE  optical
($V,I$) data for ACs to construct accurate optical $P$--$L$,
$P$--$L$--$C$ and $P$--$W$ relations both for F-
and FO-mode ACs. These relations were obtained for the first time
from a sample of pulsators covering in a uniform and complete way
the AC instability strip. 

The application of the $P$--$W(V,I)$ relation to three dwarf 
galaxies hosting significant populations of ACs,
revealed that this relation is a valuable tool for deriving distances
within the Local Group. Due to the lower dispersion, we expect that
the  $P$--$W(V,K_\mathrm{s})$ first derived in this paper  will become 
an even better tool for measuring distances  to galaxies hosting
ACs. More NIR ($K_\mathrm{s}$-band in particular) data for ACs  
in other Local Group galaxies are needed to properly exploit the
properties of the $P$--$W(V, K_\mathrm{s})$  relation. 
%
%

\section*{Acknowledgments}

We thank our anonymous Referee for his/her very helpful comments 
that helped in improving the paper. 
V.R. warmly thanks Roberto Molinaro for providing the
program for the spline interpolation of the light curves.
We thank K. Bekki and R. Guandalini for helpful discussions. 
Partial financial support for this work was provided by PRIN-INAF 2011
(P.I. Marcella Marconi) and PRIN MIUR 2011 (P.I. F. Matteucci).
We thank the UK's VISTA Data Flow System comprising the VISTA pipeline
at the Cambridge Astronomy Survey Unit (CASU) and the VISTA Science
Archive at Wide Field Astronomy Unit (Edinburgh) (WFAU) 
for providing calibrated data products supported by the STFC.
RdG acknowledges research support from the National
Natural Science Foundation of China (NSFC) through grant 11073001.




\end{document}